\documentclass[fleqn,10pt,twocolumn]{wlscirep}
\usepackage[utf8]{inputenc}
\usepackage[T1]{fontenc}
\newcommand{\fl}[1]{\textcolor{blue}{\textbf{#1}}}
\usepackage{siunitx}

\title{On-chip interference of scattering from two individual molecules in plastic}

\author[1,2,+]{Dominik Rattenbacher}
\author[1,+]{Alexey Shkarin}
\author[1]{Jan Renger}
\author[1]{Tobias Utikal}
\author[2,1,3]{Stephan G\"otzinger}
\author[1,2,*]{Vahid Sandoghdar}

\affil[1]{Max Planck Institute for the Science of Light, Erlangen, D-91058 Erlangen, Germany}

\affil[2]{Friedrich-Alexander-Universit\"at Erlangen-N\"urnberg,
Department of Physics, D-91058, Erlangen, Germany}

\affil[3]{Friedrich-Alexander-Universit\"at Erlangen-N\"urnberg,
Graduate School in Advanced Optical Technologies (SAOT), D-91058, Erlangen, Germany}

\affil[*]{vahid.sandoghdar@mpl.mpg.de}

\affil[+]{these authors contributed equally to this work}

\begin{abstract}
\textbf{Integrated photonic circuits offer a promising route for studying coherent cooperative effects of a controlled collection of quantum emitters. However, spectral inhomogeneities, decoherence and material incompatibilities in the solid state make this a nontrivial task. Here, we demonstrate efficient coupling of a pair of organic molecules embedded in a plastic film to a $\mathbf{TiO}_2$ microdisc resonator on a glass chip. Moreover, we tune the resonance frequencies of the molecules with respect to that of the microresonator by employing nanofabricated electrodes. For two molecules separated by a distance of about 8$\,$µm and an optical phase difference of about $\pi/2$, we report on a large collective extinction of the incident light in the forward direction and the destructive interference of its scattering in the backward direction. Our work sets the ground for the coherent coupling of several molecules via a common mode and the realization of polymer-based hybrid quantum photonic circuits.}

\end{abstract}
\begin{document}
\flushbottom
\maketitle

Optical microcavities have been employed in a variety of arrangements for manipulating and enhancing the coupling of light with single quantum emitters \cite{Haroche2006,Pscherer2021,Tomm2021,Niemietz2021}. The emerging efforts for quantum information processing have invoked the use of a large number of such composite emitter-cavity entities as nodes for building quantum networks \cite{Kimble2008}. Indeed, there have been some reports on the coupling of two individual emitters in the near field \cite{Hettich2002, Unold2005, Trebbia2022} or via propagating modes in free space \cite{DeVoe1996,Lettow2010,Rezus2011,Ritter2012,Schug2013,Hermans2022}. However, upscaling the number of emitters will only be feasible in integrated photonic circuits, similar to the recent advances in circuit quantum electrodynamics (QED) \cite{Majer2007,Loo2013,Mirhosseini2019,Kannan2023}. Quantum dots and color centers are currently among the most popular solid-state platforms for this purpose \cite{Evans2018, Tiranov2022, Lukin2023}, but these systems confront many challenges posed by their matrix materials. To overcome these difficulties and limitations, hybrid arrangements combining various emitters and photonics circuits would be highly desirable \cite{Sandoghdar2020,Wan2020}. 

Polycyclic aromatic hydrocarbon (PAH) molecules present an attractive choice for a hybrid solution as they offer excellent optical properties, high doping concentrations, many wavelength options, DC-Stark frequency control and a variety of host materials with a range of fabrication methods \cite{Toninelli2021}. Single PAHs have been successfully coupled to ${\rm TiO}_2$, Si$_3$N$_4$ and GaP waveguides \cite{Tuerschmann2017,Lombardi2018,Shkarin2021,Boissier2021,Ren2022} and micro-ring resonators \cite{Rattenbacher2019}. In these efforts, the emitters were doped in an organic crystal confined to a micro-channel \cite{Tuerschmann2017,Shkarin2021,Boissier2021} or in a microcrystal that was carefully placed in the vicinity of a waveguide \cite{Lombardi2018,Ren2022}. However, scattering at crystal boundaries leads to increased propagation losses \cite{Rattenbacher2019}. To circumvent this issue, here we turn to polymers as host matrix since they can form smooth coverages of the complex topologies of an integrated circuit. Indeed, we have verified that the finesse of micro-ring resonators improves by a factor of 40 when covered by polymethyl methacrylate (PMMA) instead of an organic crystal.

\begin{figure*}[t!]
\centering
	\includegraphics[width=0.9\textwidth]{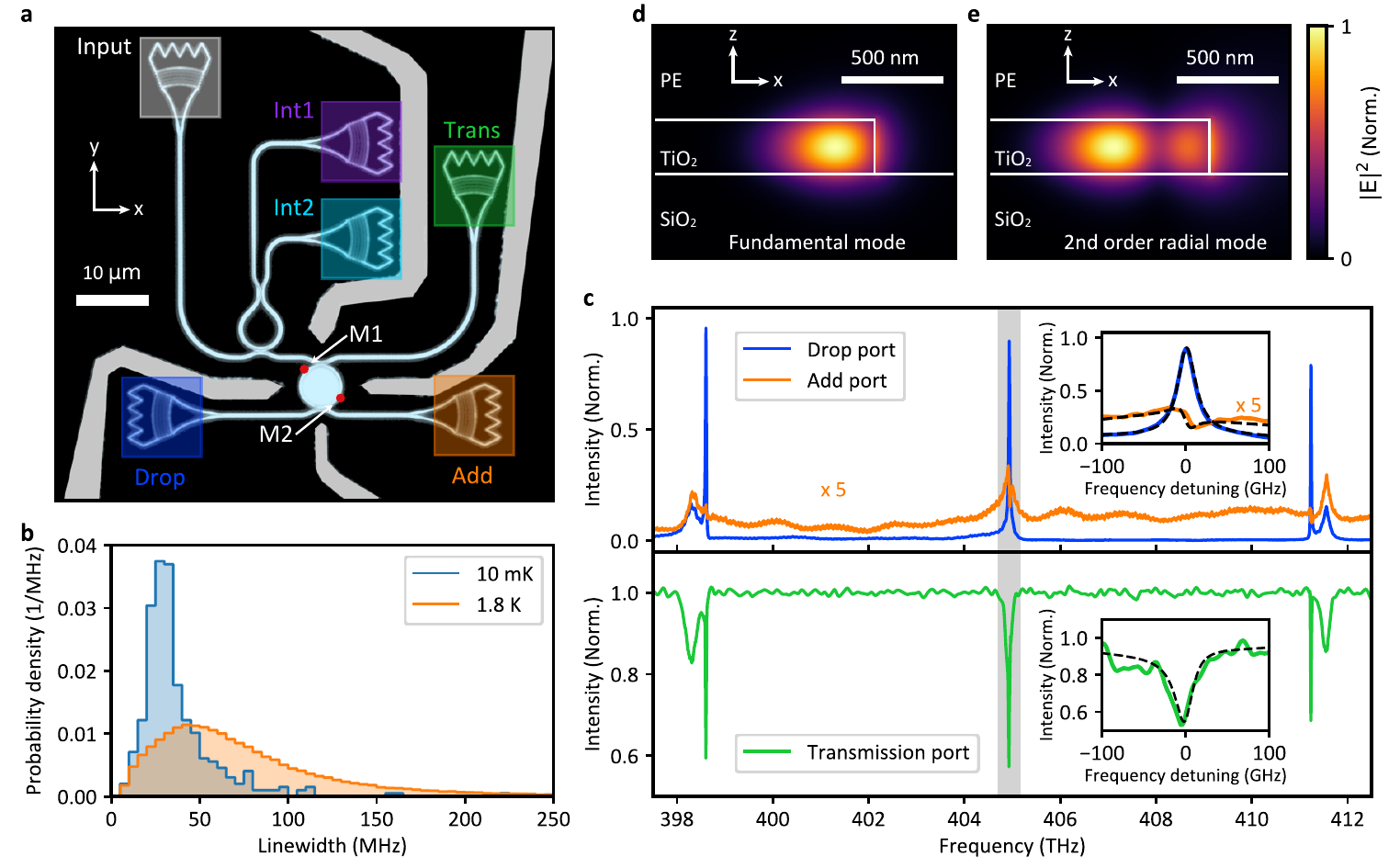}
	\caption{\textbf{a}, Micrograph of the circuit used in this work. The modes of the disc microresonator can be excited and read out via two waveguides whose ports are color-coded: white (input), blue (drop port), orange (add port), green (transmission port), and purple, turquoise (interferometer ports). The two red dots on the disc resonator indicate the two molecules, M1 and M2, used in this work. The electrodes are shown in light grey. We applied a voltage difference $\Delta V$ between the lower left pair and the upper right pair.
	\textbf{b}, Linewidth distribution for DBT in PE at a temperature of \SI{10}{\milli\kelvin} (blue) and \SI{1.8}{\kelvin} (orange). \textbf{c}, Spectra of the resonator at the drop (blue), add (orange) and transmission (green) ports. The insets show close-ups of the central resonance at \SI{404.935}{THz}, highlighted by the gray rectangle. Black dashed lines show fits according to the theory model discussed later in the text. \textbf{d},\textbf{e}, Simulated intensity profiles of the fundamental and the second-order radial modes of the disc. The modes are plotted in the x-z plane, i.e., the plane orthogonal to the  micrograph in \textbf{a}. The refractive indices used in the simulations were 1.45 for SiO$_2$, 1.52 for PE and 2.34 for TiO$_2$.
	\label{fig:resonator}}
\end{figure*}

\subsection*{Results}

An optical micrograph of our sample is displayed in Fig.~\ref{fig:resonator}\fl{a}. We etched a disc resonator (diameter \SI{6}{\micro\meter}) out of a \SI{255}{\nano\meter} thick $\mathrm{TiO}_2$ layer obtained by atomic layer deposition on a $\mathrm{SiO}_2$ substrate. The resonator was interfaced with two waveguides (width \SI{250}{\nano\meter}) terminated with grating couplers for efficient free-space connectivity. To optimize the evanescent coupling between the disc and the waveguides, the latter were curved with a radius of \SI{3}{\micro\meter} next to the disc. An integrated interferometer provides means for analyzing the phase of the light circulating in the counter-clockwise (CCW) mode of the disc. Furthermore, four chromium microelectrodes were fabricated close to the resonator to apply local DC fields. The circuit was then covered by polyethylene (PE), which is a common plastic with widespread use in the packaging industry. PE can result in homogeneous and optically transparent films \cite{Wirtz2006}. 

The use of a polymer medium in a quantum optical setting might raise questions since the disordered arrangement of polymer chains is known to cause PAHs to undergo substantial dephasing and spectral diffusion, resulting in GHz linewidths \cite{Walser2009}. Indeed, PE is the only polymer that has been reported to host lifetime-limited transitions for a molecule of the PAH family (terrylene), albeit at $T=\SI{30}{mK}$ \cite{Donley2000}. This superior behavior, which has been attributed to local crystallization of PE \cite{Orrit1992,Wirtz2006}, motivated us to use PE. In our work, we doped PE with another PAH molecule, namely dibenzoterrylene (DBT) \cite{Nicolet2007b,Verhart2016,Ren2022} by submerging the film in a 100\,ppm solution of DBT/trichlorobenzene.

The blue-shaded histogram in Fig.~\ref{fig:resonator}\fl{b} displays the linewidth distribution of the zero-phonon lines (ZPL) measured at $T=\SI{10}{mK}$ in a dilution refrigerator. It is sharply peaked around the expected natural linewidth of \SI{30}{MHz} \cite{Nicolet2007b,Verhart2016}, similar to the results for terrylene in PE \cite{Donley2000}. 
Unlike terrylene, however, the orange-shaded histogram in Fig.~\ref{fig:resonator}\fl{b} shows that a considerable fraction of DBT molecules reaches the natural linewidth at a more accessible temperature of \SI{1.8}{K}, with the most probable linewidth amounting to \SI{45}{MHz}. The molecular resonances do experience rare and discrete photo-induced spectral jumps between several frequencies. However, this process can be reversed either by locally exciting the molecule at its new resonance frequency or by thermo-cycling the sample up to \SI{20}{K}. In practice, we could work with the same individual molecules for several months despite occasional jumps. Given the narrow linewidths and stable transitions in DBT:PE, we performed the remaining experiments of this paper in a helium bath cryostat at \SI{1.8}{K}. 

Figure~\ref{fig:resonator}\fl{c} presents a spectrum of the microdisc resonator covered by PE recorded at the different ports. The insets display a close-up of the central resonance at \SI{404.935}{THz}, which was used in this work. The spectra show that the dominant mode exhibits a full width at half-maximum (FWHM) value of $\kappa=2\pi\cdot$\SI{27}{GHz} and a free spectral range (FSR) of about 6.3\,THz (\SI{11.6}{nm}), corresponding to a finesse of 230 and a Q-factor of 15,000. The resonator is under-coupled to the waveguides, resulting in a transmission dip of $\approx\;$\SI{40}{\percent}. We also note a second resonance with a slightly larger FSR and much broader FWHM, which we attribute to the second-order radial mode. Figures~\ref{fig:resonator}\fl{d} and \ref{fig:resonator}\fl{e} portray the simulated mode profiles for the TE-like fundamental and second-order radial modes. The molecules presented in the following are predominantly coupled to the narrower fundamental mode, but we shall also consider the second mode for a quantitative description of our system. We remark that the spectral overlap of the two modes in the region of interest of this study was coincidental. 

\begin{figure}[t!]
\centering
	\includegraphics[width=0.48\textwidth]{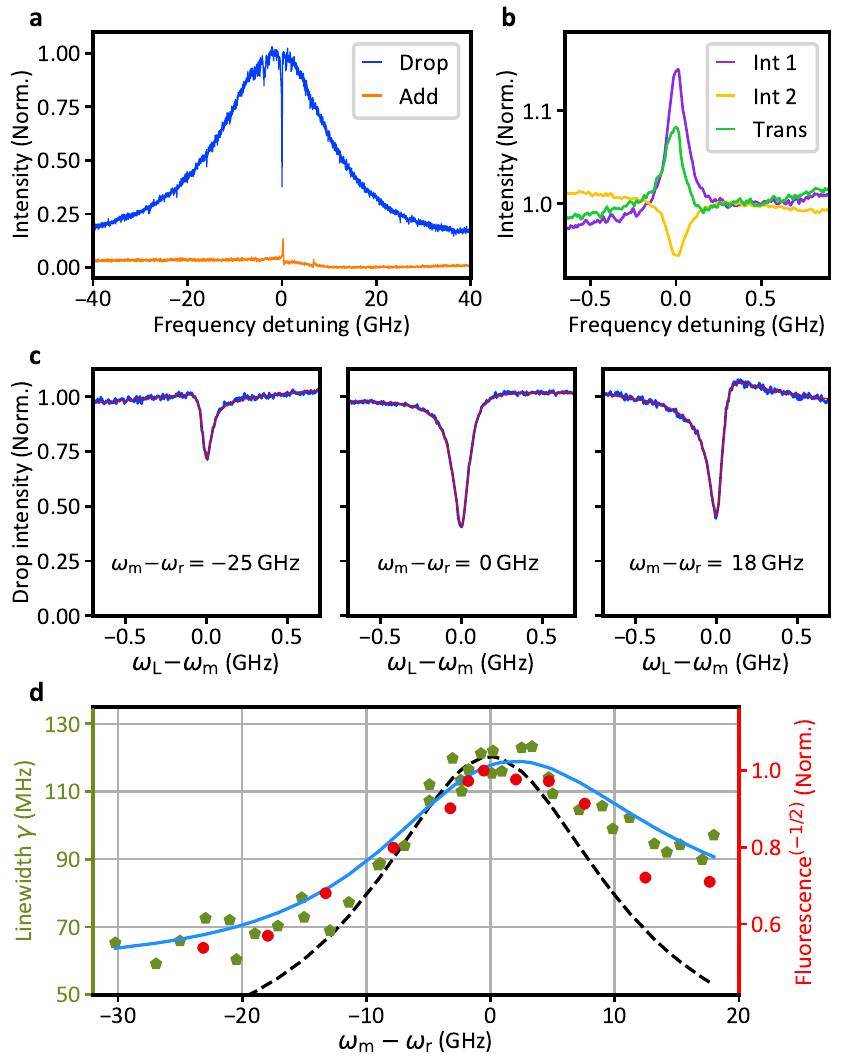}
	\caption{\textbf{a}, Drop and add port spectra for a single well-coupled molecule resonant with the microdisc. \textbf{b}, Scans of the same molecule in the interferometer ports (purple, turquoise) and in the transmission port (green).  \textbf{c}, Drop port spectra of the same molecule for different frequency detunings from the resonator as indicated in the legend. Red curves show complex Lorentzian fits. \textbf{d}, Linewidths $\gamma$ (green) extracted from the fits as a function of the detuning from the resonator. The blue curve presents a theoretical fit according to our input-output model, taking into account the fundamental and second-order radial modes (see text). The black dashed curve is the outcome of a theoretical treatment considering only the fundamental degenerate pair of the resonator modes. The red symbols plot the inverse square root of the molecule's fluorescence signal measured in a confocal excitation scheme (see text).
	\label{fig:single_molecule}}
\end{figure}

To achieve a match between the cavity spectrum and the inhomogeneous band of DBT, which spans over a few THz, we fabricated a series of discs with increasing diameters in steps of \SI{100}{nm} on each chip, corresponding to frequency increments of about \SI{0.9}{THz}. The drop port signal (blue) in Fig.~\ref{fig:single_molecule}\fl{a} displays the resonance of a single molecule (M1) as a dip on a microdisc resonance. The orange curve in this figure shows a peak caused by the molecule scattering light into the CCW mode, which is detected at the add port. One can also identify other weakly-coupled molecules. The measurements were taken in the weak-excitation regime at a saturation parameter of $S=0.02$. Figure~\ref{fig:single_molecule}\fl{b} presents a close-up of the dominant molecule signal in transmission and in the two interferometer ports. The exact shapes of the molecular resonance at the interferometer ports depend on the position of the molecule along the disc circumference. By using an aspheric lens, we could also excite and image individual molecules in the direction normal to the chip plane, thus locating them with respect to the microdisc with subwavelength precision. The position of the molecule presented in Fig.~\ref{fig:single_molecule} is marked as M1 in Fig.~\ref{fig:resonator}\fl{a}.

The spectral features of a coupled molecule depend on the detuning of its transition frequency ($\omega_\mathrm{m}$) from the resonator center frequency ($\omega_\mathrm{r}$). The blue spectra in Fig.~\ref{fig:single_molecule}\fl{c}  present three examples of a single-molecule resonance as we Stark-shifted $\omega_\mathrm{m}$ by applying up to 200\,V across the integrated microelectrodes \cite{Lettow2010,Moradi2019,Shkarin2021}. The red curves display the outcome of fitting a complex Fano line shape to the experimental spectra according to 
\begin{equation}
    f(\omega_{\rm L})=f_\mathrm{r}(\omega_{\rm L}-\omega_{\rm r})\left\vert 1-\frac{ A \mathrm{e}^{\mathrm{i}\phi}}{1+\mathrm{i}\frac{2(\omega_{\rm L}-\omega_\mathrm{m})}{\gamma}}\right\vert^2,
\end{equation}
where $\omega_{\rm L}$ denotes the laser frequency and $f_\mathrm{r}(\omega)$ is a real function that describes the resonator profile in the absence of the molecule. Moreover, $\phi$, $\gamma$ and $A$ stand for the phase, FWHM, and relative amplitude of the molecular signal, respectively. On resonance, this molecule demonstrates a Purcell-enhanced linewidth of $2\pi\cdot$\SI{125}{MHz}. Figure~\ref{fig:single_molecule}\fl{d} shows the recorded values of $\gamma$ (green symbols) as a function of frequency detuning. A simple model based on the fundamental CW and CCW radial modes predicts that the increase in $\gamma$ should be proportional to the cavity resonance profile (black dashed line), but the experimental data in Fig.~\ref{fig:single_molecule}\fl{d} indicate a clear deviation. Hence, we derived an extended input-output model \cite{Collett1984,Gardiner1985,Auffeves2007} accounting for the presence of the second-order radial mode and a weak backscattering that mixes the CW and CCW modes \cite{Srinivasan2007,Srinivasan2007b,Mazzei2007}. A fit based on this model yields a good agreement with the measured linewidth of the molecular resonance (see blue curve in Fig.~\ref{fig:single_molecule}\fl{d}). Extrapolating the model lets us deduce a molecular linewidth of $\gamma^0=2\pi\cdot$\SI{33}{MHz} far from the cavity resonance, consistent with the known value of the natural linewidth for DBT in crystalline hosts \cite{Nicolet2007b,Verhart2016} and with the peak of the linewidth distribution in PE at $T=\SI{10}{mK}$ (see Fig.~\ref{fig:resonator}\fl{b}).

The linewidth $\gamma$ of the molecular resonance has two main contributions stemming from the decay via the ZPL ($\gamma_\mathrm{zpl}$) and red-shifted fluorescence to higher-lying vibrational levels of the electronic ground state ($\gamma_\mathrm{red}$). In other words, $\gamma=\gamma_\mathrm{zpl}+\gamma_\mathrm{red}$. The parameter $\alpha=\gamma_\mathrm{zpl}/\gamma$ defines a branching ratio as the fraction of the molecular emission via the ZPL. When coupled to a resonator, the ZPL transition is enhanced by the Purcell factor $F$ such that $\gamma^\prime=(1+F)\gamma^0_\mathrm{zpl}+\gamma^0_\mathrm{red}$\cite{Wang2019}, where the prime and $0$ superscripts denote the Purcell-enhanced and uncoupled cases, respectively. Assuming the same free-space branching ratio $\alpha^0=1/3$ for DBT:PE as for DBT:Ac \cite{Nicolet2007b}, we obtain $F\approx9$ and $\alpha^\prime=\,$ \SI{82}{\percent} for M1. 

In Fig.~\ref{fig:single_molecule}\fl{d}, we also present the result of fluorescence measurements as a function of the cavity detuning when the molecule was excited and detected confocally via the out-of-plane optical access. The overlap of the red symbols and the linewidth data (green) shows that the fluorescence signal scales with $1/\gamma^{2}$. Two phenomena contribute to this behavior. First, the excitation becomes less efficient for larger linewidths. Second, once the molecule is excited, the relative probability to emit a red-shifted fluorescence photon is reduced for larger $\alpha$. Since each effect scales as $1/\gamma$, their combination leads to the observed trend.

To assess the observed extinction dip, it is instructive to commence by considering a beam of light traversing through a dipole in free space. In this case, the incident light can be perfectly extinguished if its spatial mode matches that of a radiating dipole \cite{Zumofen2008}. Similarly, when a molecule is placed in a Fabry-Perot cavity, the degree of overlap between the cavity mode and its emission given by $\beta=F\gamma^0_\mathrm{zpl}/\gamma$ determines the extinction dip \cite{Auffeves2007, Wang2019}, whereby the fraction of light that is transmitted at the molecular resonance is given by $(1-\beta)^2$. The situation is different in a ring resonator, where one has to account for two degenerate modes (CW and CCW) which couple equally strongly to the emitter \cite{Weiss1995,Srinivasan2007,Srinivasan2007b,Mazzei2007}, resulting in a coupling of $\beta/2$ per mode. In this case, the extinction of the drop port corresponds to a transmission of $(1-\beta/2)^2$, reaching \SI{25}{\percent} for a perfectly coupled emitter ($\beta$=1).

For M1, the measured extinction dip amounts to \SI{61}{\percent}, corresponding to $\beta\approx 0.75$ and QED parameters $\{g,\kappa,\gamma^0\}=2\pi \cdot\{0.49,27,0.03\}\,\mathrm{GHz}$, where $g$ denotes the molecule-resonator mode coupling strength. Given the large $\beta$, the molecule efficiently scatters the CW incident field into the CCW mode, generating a standing wave with a visibility of about \SI{89}{\percent} and a node at the position of the molecule \cite{Srinivasan2007}. The underlying physics of this phenomenon is similar to that of a classical scatterer that leads to mode splitting in whispering-gallery resonators \cite{Weiss1995,Srinivasan2007b,Mazzei2007}. However, the difference in the scattering phases between a dielectric nano-scatterer ($\sim 0 \pi$) and a resonant quantum emitter ($\pi/2$) results in important implications for the spectrum of the system. 


\begin{figure*}[ht!]
\centering
	\includegraphics[width=0.95\textwidth]{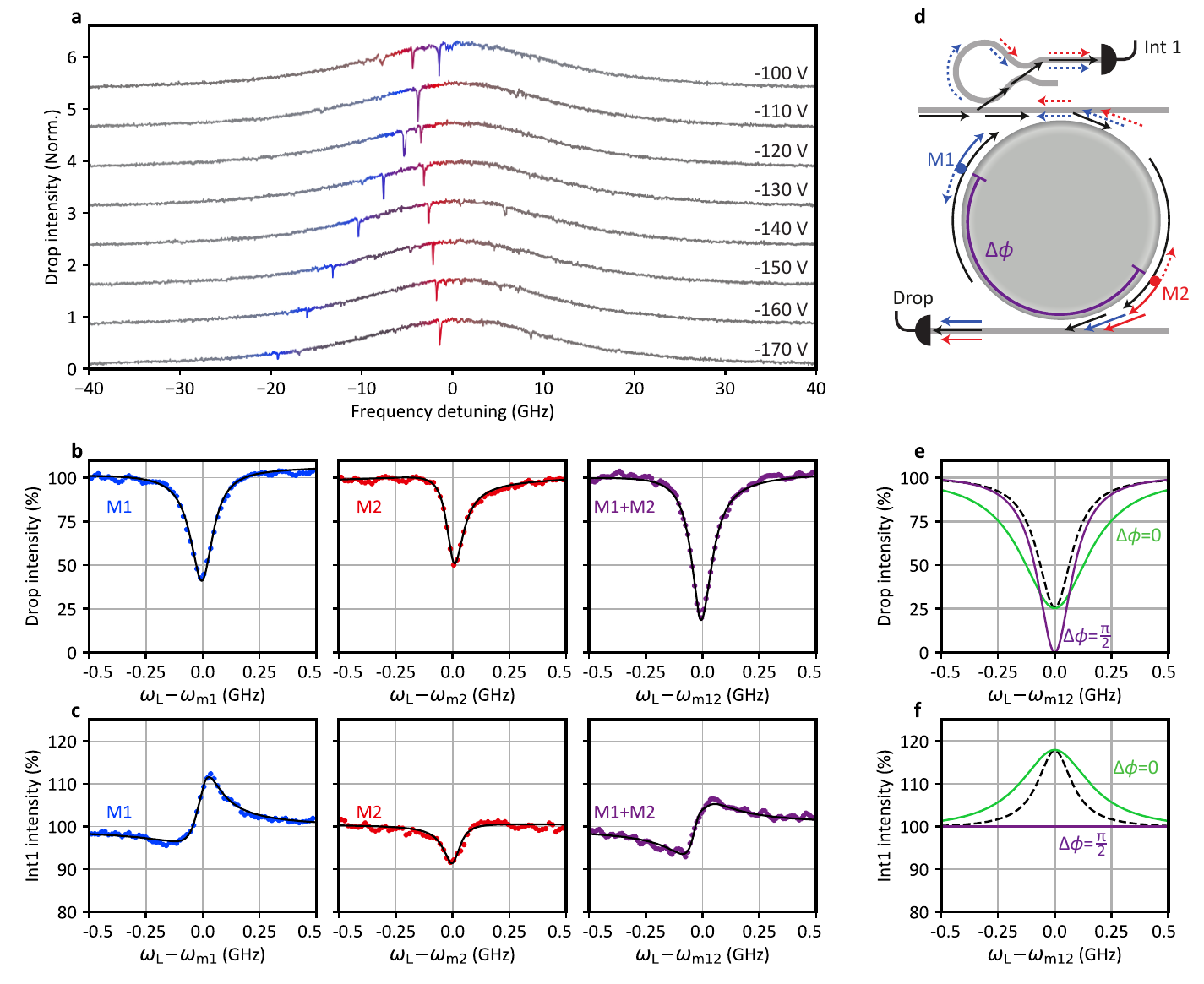}
	\caption{\textbf{a}, Waterfall plot of the drop port spectrum for several applied voltages. The two well-coupled molecules are highlighted in red and blue. The applied external voltage is indicated above each curve. \textbf{b},\textbf{c}, Drop port and interferometer port 1 signals of each individual molecule (M1: blue, M2: red) and of the two molecules in resonance (violet) versus the laser detuning from the molecular resonance. For the violet spectrum, the detuning from the mean molecule frequency $\omega_{m12}=(\omega_{m1}+\omega_{m2})/2$ is shown. All spectra were recorded at a detuning of about \SI{4}{GHz} from the center of the microdisc resonance. Black curves are fits according to the input-output model. \textbf{d}, Sketch of the detection scheme and the optical paths in the waveguides (solid line: CW, dashed line: CCW, black: excitation, blue/red: light scattered by M1/M2). \textbf{e},\textbf{f} Theoretical spectra of an individual molecule (black dashed lines) and two resonant molecules at the drop port and interferometer port 1 for phase differences $\Delta\phi=0$ (green) and $\Delta\phi=\pi/2$ (violet). The coupling efficiency is set to $\beta=$\SI{99}{\percent} for both molecules.
	\label{fig:two_molecules}}
\end{figure*}
We now introduce a second molecule (M2) that is also efficiently coupled to the resonator and is located on the opposite side of the disc with respect to M1 (see Fig.~\ref{fig:resonator}\fl{a}). By applying an electric potential $\Delta V$ between two pairs of microelectrodes, we tune the ZPL frequencies of M1 and M2 into resonance. As seen in Fig.~\ref{fig:two_molecules}\fl{a}, the two molecules become resonant at a voltage of \SI{-110}{V} and a detuning of \SI{-4}{GHz} from the resonator frequency.  The data in Fig.~\ref{fig:two_molecules}\fl{b} show the spectra of M1 (blue) and M2 (red) when the other molecule was detuned by many molecular linewidths. The extinction dips of about \SI{61}{\percent} and \SI{49}{\percent} correspond to $\beta_1=$\SI{75}{\percent}, $g_1=$\SI{490}{MHz}, $\beta_2=$\SI{56}{\percent} and $g_2=$\SI{395}{MHz} for M1 and M2, respectively. The violet symbols in Fig.~\ref{fig:two_molecules}\fl{b} report an extinction dip of about \SI{81}{\percent} at the drop port when the two molecules were tuned to the same frequency. The observed spectrum results from the interference between coherently scattered light of the two molecules. To assess the phase difference between the two fields, we used the integrated interferometer (see Fig.\,\ref{fig:resonator}\fl{a}). The peak observed for M1 (red data, Fig.~\ref{fig:two_molecules}\fl{c}) and the dip observed for M2 (blue data, Fig.~\ref{fig:two_molecules}\fl{c}) at the interferometer port 1 indicate that the back-scattered fields by the two molecules are out of phase. Consequently, their signals partially cancel, as shown by the violet spectrum in Fig.~\ref{fig:two_molecules}\fl{c}.

We analyzed the data presented in Fig.~\ref{fig:two_molecules}\fl{b,c} by extending the input-output formalism to account for the second emitter. Here, we first fitted the drop and interferometer port spectra of the individual molecules (solid black lines, Fig.~\ref{fig:two_molecules}\fl{b,c}), obtaining a phase difference of $\Delta\phi=0.58\pi$ between their fundamental mode coupling constants. We note that, as illustrated in Fig.~\ref{fig:two_molecules}\fl{d}, the components of the CW incident light that are scattered by the two molecules into the CCW direction accumulate twice the phase difference. Using the properties of M1 and M2 obtained from their individual fits, we obtain an excellent agreement between the resonant data and the model for a residual frequency detuning of $\omega_{\rm m1}-\omega_{\rm m2}=2\pi\cdot\SI{10}{MHz}$ (solid black curves, right panel of Fig.~\ref{fig:two_molecules}\fl{b,c}). The measured phase difference of $2\Delta \phi \sim \pi$ implies that the back-scattered fields of the two molecules interfere destructively in the CCW mode. In other words, the light emitted by the two molecules couples mainly to the CW mode. This is, indeed, confirmed by the strong extinction dip of \SI{81}{\percent} at the drop port, exceeding the maximum value of \SI{75}{\percent} expected for a single molecule with $\beta=1$. Correspondingly, the back-scattered two-molecule signal at the interferometer port, which is approximately the sum of the individual signals from M1 and M2, is reduced. Indeed, for two identical emitters with $2\Delta \phi=\pi$, one would expect a flat interferometer signal as depicted by the violet curves in Fig.~\ref{fig:two_molecules}\fl{e,f}. Theory curves for a single molecule are plotted as black dashed lines for reference. Suppression of one of the two counter-propagating modes in a ring cavity has recently been discussed as a chirality phenomenon and was demonstrated in the context of spin-orbit coupling \cite{ Lodahl2017b,Cano2019}. Our data suggest an alternative approach based on interference, as recently also reported in circuit QED \cite{Kannan2023}. 

\subsection*{Discussion and Outlook}

We demonstrated a hybrid photonic circuit based on $\mathrm{TiO}_2$ elements on a glass chip covered by a plastic film containing organic dye molecules. The use of a polymer coating and shrinking the microresonators down to a diameter of \SI{6}{\micro\meter} allowed us to achieve a 14-fold increase in finesse as compared to our previous work \cite{Rattenbacher2019}. Although lack of long-range order in polymer matrices has been usually associated with dephasing and spectral jumps \cite{Walser2009}, the combination of DBT and PE exhibited narrow linewidths, with a significant fraction of molecules reaching their lifetime limit in a helium bath cryostat. These results pave the way for the use of formable plastic material in quantum photonics devices \cite{Schell2013,Colautti2020b,Siena2022}, possibly also in combination with other systems such as semiconductor quantum dots \cite{Lee2000,Labeau2003,Xia2021}. Indeed, the higher the Purcell factor, the more one could afford to work with a certain degree of intrinsic dephasing. 

Currently, the main challenge for reaching higher Purcell factors is the broadband absorption of weakly coupled and spectrally broadened DBT molecules. This might be ameliorated by a further annealing step on the doped PE film, selective bleaching of the background molecules, or by lowering the doping concentration. Indeed, we found that the finesse of discs in pristine PE films is about 4 times higher than in doped ones, an improvement that would suffice to reach $\beta \sim 92\%$ in our system. Another option for reaching larger coupling efficiencies would be to deliver molecules locally to certain positions of the structure via nanoprinting \cite{Hail19} or pick-and-placement of nanocrystals \cite{Pazzagli2018,Lombardi2018}. 

By using microelectrodes and an interferometer on the chip, we tuned the resonances of two molecules located at a large distance of several micrometers and examined their interplay. For molecule spacings corresponding to an optical phase difference of $\Delta\phi=\pi/2\,(\mathrm{mod}\:\pi)$, one enhances the extinction of a laser beam and suppresses the backward scattering (violet curves, Fig.~\ref{fig:two_molecules}\fl{e,f}), leading to directional selectivity \cite{Kannan2023,Lukin2023}. We note that the yield for identifying two molecules that efficiently coupled to the resonance of a microdisc was low in our current experiment. However, by applying automated selection protocols to a large number of nanofabricated resonators, it will be possible to address this issue.

A phase difference of $\Delta\phi=0\,(\mathrm{mod}\:\pi)$ between the two molecules in our geometry leads to a constructive interference in both CW and CCW directions. To analyze this arrangement, it is convenient to switch from the propagating resonator mode basis ($\hat{a}$, $\hat{b}$) to the standing wave basis $\hat{S}_\pm=\frac{1}{\sqrt{2}}(\hat{a}\pm\hat{b})$. In this frame, the molecules are only coupled to one standing wave, and the system becomes analogous to the case of two emitters in a Fabry-Perot cavity with effective single-emitter coupling strength of $g_\mathrm{eff}=\sqrt{2}g$. The interaction of the two emitters mediated by the resonator mode leads to the hybridization of their energy levels, giving rise to super- and subradiant states. The super-radiant state experiences twice the Purcell effect of a single molecule and undergoes $\sqrt{2}$ enhanced coupling efficiency while the resulting subradiant state is decoupled from the cavity and exhibits the natural linewidth of an individual molecule. Figure~\ref{fig:two_molecules}\fl{e} compares the extinction spectra of one emitter (black) and two emitters (green) with $\beta=$\SI{99}{\percent}. It is to bear in mind that the hybridized ``super-molecule'' is still coupled to two counter-propagating modes so that its extinction dip does not surpass \SI{75}{\percent}.

In this work, we reached $\beta\approx$\SI{75}{\percent}, $F=9$, and cooperativity $C=\frac{4g_\mathrm{eff}^2}{\kappa\gamma^0}\approx3$. These parameters would lead to super- and sub-radiant linewidths of \SI{210}{MHz} and \SI{30}{MHz}, respectively, for two resonant molecules with $\Delta\phi=0\,(\mathrm{mod}\:\pi)$. Detuning the emitter pair from the cavity resonance by one cavity linewidth $\kappa$ would result in super- and sub-radiant linewidths of \SI{66}{MHz} and \SI{30}{MHz}, respectively, and an exchange interaction strength of $J=\frac{g_\mathrm{eff}^2}{\kappa}=$\SI{20}{MHz}. Our experimental architecture makes it possible to scale up the number of photon-mediated emitters distanced by much larger than a wavelength, thus, realizing an optical paradigm similar to the achievements in circuit QED \cite{Loo2013,Mirhosseini2019}. Such experiments will also provide a foundational approach to the understanding of the recent results in room-temperature strong coupling of large emitter ensembles \cite{Ebbesen2016,Genet2021}. 
 
\subsection*{Methods}
\subsubsection*{Chip fabrication}

The fabrication of the nanophotonic circuit starts with the preparation of a homogeneous ${\mathrm{TiO}}_{2}$ layer on a fused silica substrate via atomic layer deposition. Using ellipsometry, we determined a layer thickness of \SI{255}{nm} and a refractive index of $n_\mathrm{TiO_2}=2.34$ at a wavelength of \SI{750}{\nano\meter} for the sample studied in this work. Next, the sample is covered by a chromium layer which acts as a hard mask. We pattern the photonic circuit by electron beam lithography employing a negative-tone resist (AR-N7520, Allresist GmbH). The pattern transfer into the Cr hard mask is done using chlorine-based reactive ion etching. The ${\mathrm{TiO}}_{2}$, which is not protected by the hard mask, is then removed with an $\mathrm{SF}_6$/$\mathrm{CF}_4$ plasma etch. After etching the structure, the mask is removed by wet chemical etching. The micro-electrodes are added in a second step. We pattern the electrode layout into a positive-tone photoresist (AZ MIR 701, Microchemicals GmbH) using direct laser writing. Then the 100-nm-thick chromium electrode is deposited and the excess material is lifted off. Finally, the electrodes are connected to copper wires using a conductive epoxy glue (H20E, Epotek).

\subsubsection*{PE film preparation and molecule doping}

Due to its high chemical stability, polyethylene (PE) can only be solved at elevated temperatures of $\gtrapprox$\;\SI{100}{\degreeCelsius} when using common solvents such as decalin, toluene and 1,2,4-trichlorobenzene (TCB). A previously published recipe for producing optically transparent PE films utilized spin coating at elevated temperatures \cite{Wirtz2006}. However, we found that the film quality critically depends on the temperature during spin coating, which is difficult to control. Therefore we developed a flow coating approach where a drop of polymer solution is blown over the sample at a defined temperature on a hotplate. This simplifies the process and improves the reproducibility of our film preparation.

We start with high-density PE pellets from Sigma-Aldrich (melt index \SI{2.2}{\g}/\SI{10}{\min}) and dissolve them in TCB (spectrophotometric grade, >\SI{99}{\percent} from Alpha Aesar/Thermo Fisher Scientific) at \SI{130}{\degreeCelsius} at a concentration of \SI{30}{\g\per\l}. Next, the cleaned sample is preheated on a hotplate at \SI{130}{\degreeCelsius}. A \SI{30}{\micro\l} drop of preheated (\SI{130}{\degreeCelsius}) PE:TCB solution is then deposited on the sample and allowed to distribute for \SI{10}{s}. A preheated heat gun at a distance of \SI{10}{cm} from the sample is switched on and the PE:TCB drop is slowly blown over the sample, leaving a thin film behind. TCB is subsequently evaporated during a drying time of \SI{10}{\minute} leaving behind the optically transparent thin PE film. The thickness can be tuned via the viscosity of the PE:TCB solution, either by changing the concentration or the temperature. By cutting the film with a razor blade and measuring the depth of the cut, we determined the film thickness for the above-mentioned parameters to be roughly \SI{700}{\nano\meter}. The refractive index was determined by ellipsometry to be $n_\mathrm{PE}=1.52$ at a wavelength of \SI{750}{\nano\meter}.

The PE film is doped with DBT by infusing the latter with a DBT/TCB solution. We prepare the solution by solving DBT at a concentration of \SI{150}{\micro\g\per\milli\l} in TCB. The PE-coated sample is placed in a home-made windowless desiccator and covered with \SI{40}{\micro\liter} of the DBT:TCB solution. After \SI{10}{\minute} at ambient pressure it is evacuated to \SI{0.01}{\milli\bar} and left at this pressure for about 12 hours. Afterwards the sample is inserted into the cryostat.\\

\textit{Acknowledgment.} We acknowledge financial support by the Max Planck Society, the RouTe Project (13N14839) through the German Federal Ministry of Education and Research (BMBF) and TRR 306 (QuCoLiMa) through the German Research Foundation (DFG). A.S. acknowledges support from an Alexander von Humboldt fellowship.

\textit{Author contributions.}
D.R. and A.S. performed the experiments and analyzed the data. J.R. fabricated the photonic chip. T.U., S.G. and V.S. supervised the project. The manuscript was written by D.R., A.S. and V.S. and all authors commented on it.


\end{document}